\documentclass[conference]{IEEEtran}
\IEEEoverridecommandlockouts
\usepackage{cite}
\usepackage{amsmath,amssymb,amsfonts}
\usepackage{graphicx}
\usepackage{subfigure}
\usepackage{textcomp, color}
\usepackage{tikz}
\usetikzlibrary{decorations.pathreplacing}
\usetikzlibrary{shapes}
\usepackage[algoruled, linesnumbered]{algorithm2e}

\def\BibTeX{{\rm B\kern-.05em{\sc i\kern-.025em b}\kern-.08em
    T\kern-.1667em\lower.7ex\hbox{E}\kern-.125emX}}

\usepackage{etoolbox}
\makeatletter
\patchcmd{\@maketitle}
{\addvspace{0.5\baselineskip}\egroup}
{\addvspace{-0.7\baselineskip}\egroup}
{}
{}
\makeatother

\begin{document}
\title{\huge{Common Language for Goal-Oriented Semantic Communications: A Curriculum Learning Framework}\vspace{-4mm}\\
\thanks{This work was supported by the Office of Naval Research (ONR) under MURI Grant N00014-19-1-2621.}
}
\author{\small{Mohammad Karimzadeh Farshbafan$^\dagger$, Walid Saad$^\dagger$, and Merouane Debbah$^+$} \\
$^\dagger$ Wireless@VT, Bradley Department of Electrical and Computer Engineering, Virginia Tech, Blacksburg, VA, USA, \\
$^+$ Technology Innovation Institute, Abu Dhabi, United Arab Emirates, and \\
Mohamed Bin Zayed University of Artificial Intelligence, 9639 Masdar City, Abu Dhabi, United Arab Emirates, \\
emails: $^\dagger$ $\{$mkarimzadeh, walids$\}$@vt.edu, $^+$ $$merouane.debbah$$@tii.ae}
\maketitle
\begin{abstract} 
Semantic communications will play a critical role in enabling goal-oriented services over next-generation wireless systems. However, most prior art in this domain is restricted to specific applications (e.g., text or image), and it does not enable goal-oriented communications in which the effectiveness of the transmitted information must be considered along with the semantics so as to execute a certain task. In this paper, a comprehensive semantic communications framework is proposed for enabling goal-oriented task execution. To capture the semantics between a speaker and a listener, a common language is defined using the concept of beliefs to enable the speaker to describe the environment observations to the listener. Then, an optimization problem is posed to choose the minimum set of beliefs that perfectly describes the observation while minimizing the task execution time and transmission cost. A novel top-down framework that combines curriculum learning (CL) and reinforcement learning (RL) is proposed to solve this problem. Simulation results show that the proposed CL method outperforms traditional RL in terms of convergence time, task execution time, and transmission cost during training.
\end{abstract}
\begin{IEEEkeywords}
Semantic communication, Goal-Oriented Communication, Reinforcement Learning, Curriculum Learning.
\end{IEEEkeywords}
\vspace{-2.5mm}
\section{Introduction}
\label{Section: Introduction}
\vspace{-1.7mm}
\looseness=-1
	\hspace{-0.75mm}Next-generation wireless networks must support autonomous interactions between millions of machines through \emph{goal-oriented communications} in which three different goals must be satisfied for data transmission: a) maximizing bit accuracy, b) maximizing the semantic information (meaning), and c) maximizing effectiveness, \cite{strinati20216g}. 
At the semantic level, one must precisely convey the desired meaning of transmitted messages.
In terms of effectiveness, the network must consider how the transmitted semantics could steer the system to its goals.
Therefore, current wireless networks that are designed based on bit accuracy, must be redesigned to have goal-oriented communication.
The key challenges for enabling semantic communications for goal-oriented tasks include developing semantic-oriented metrics, defining a common language between the transmitter and receiver, and transmitting only semantic information that is strictly relevant to the system's goal. Thus, to design a truly goal-oriented communication system, semantics must be considered in cohort with the system's goals.
	

The works in \cite{xie2021deep, xie2020lite, guler2018semantic, lu2021reinforcement, yun2021attention, tucker2021emergent} looked at semantic communication without considering the goal-oriented nature of the system. In \cite{xie2021deep} and \cite {xie2020lite}, the authors proposed a semantic communication frameworks for text transmission using deep learning.
The work in \cite{guler2018semantic} introduced {\color{black}a Bayesian game to minimize the end-to-end average semantic error.} In \cite{lu2021reinforcement}, the authors proposed a reinforcement learning (RL) method to capture the meaning of transmitted information by learning semantic similarities between them.
The work in \cite{yun2021attention} introduced a model for implementing semantic communications to address the reliability and latency requirements for drone networks. The work in \cite{tucker2021emergent} introduced a neural agent architecture with the capability of communication among the agents using discrete tokens. The most significant limitation of these works in \cite{xie2021deep, xie2020lite, guler2018semantic, lu2021reinforcement, yun2021attention, tucker2021emergent} is that they are focused on a specific application and, thus, their results cannot be generalized to fully-fledged semantic for goal-oriented communication.

In \cite{maarala2016semantic} and \cite{seo2021semantics}, the authors investigated the notion of reasoning for semantic communications. The work in \cite{maarala2016semantic} developed a semantic reasoning system for a realistic Internet of things (IoT) network to enable the reasoning of actionable knowledge. Meanwhile, the authors in \cite{seo2021semantics} investigated a semantic-native communication structure for extracting the most effective semantics of the transmitter for the receiver. However, the works in \cite{maarala2016semantic} and \cite{seo2021semantics} are limited in many ways. First, they do not consider the effectiveness of the transmitted information for task execution purposes.
Second, the dynamic and random effects of the environment on the task execution process were overlooked.
\emph{To the best of our knowledge, there is no existing framework that explicitly accounts for the synergies between semantic communications and goal-oriented task execution.}

\looseness=-1
The main contribution of this paper is to address this challenge by developing a novel holistic framework that enables semantic communication for task execution purposes while minimizing transmission cost and task execution time. In particular, we introduce a comprehensive model of semantic communication for task execution, which includes a speaker and a listener who wish to execute a set of tasks using a set of common language formed by beliefs. Each task is defined as a chain of multiple events that the speaker observes. The speaker is responsible for describing the events to the listener, while the listener is responsible for taking action in the task execution procedure. Then, we formulate an optimization problem to simultaneously minimize the required time and transmission cost of task execution. These objectives are achieved by finding the most abstract and perfect description of each event based on the beliefs. Due to the difficulties of solving the proposed optimization problem, we solve it using a novel top-down curriculum learning (CL) \cite{narvekar2020curriculum} framework based on RL. In each step of the proposed CL, we consider a specific RL problem, and using the output of each step; we initialize the RL problem of the next step. Finally, we evaluate the performance of the proposed method to solve the introduced optimization using task execution time and transmission cost, compared to the traditional RL and non-semantic communication. Simulation results show that the proposed method outperforms traditional RL in convergence time, task execution time, and transmission cost. Also, the proposed goal-oriented model outperforms non-semantic communication in transmission cost.

\vspace{-2mm}
\section{System~Model}
\label{Section: System model}
\vspace{-1mm}
Consider a pair of a speaker and a listener who wish to complete a set of sequential tasks. Each task is composed of multiple events observed sequentially by the speaker. Each event captures the state of the environment as perceived by the speaker at a given time. The speaker must describe each observed event to a remote listener. In response to the received information and according to the events described by the speaker, the listener must take specific actions to steer the system towards completing the ongoing task. This is a central feature of goal-oriented communications.
Since the transmitted description captures the speaker's perception of the observed event, then, initially, the transmitted information cannot perfectly describe the observed events.
The effect of the listener's action on the task execution process is seen in the next observed event. Hence, the speaker can be aware of the accuracy of its description for each event based on the next event. Examples of such systems are IoT systems performing control tasks and automotive production in factories.

We consider an episodic structure for task execution in which each episode $m$ is dedicated to executing a specific task. Let $T_m$ be the task executed in episode $m$. For each episode, we divide time into equal slots. At the beginning of time slot $n$ of episode $m$, the speaker observes a new event, $e_{m,n}^S$, and is responsible for transmitting its deduced information about $e_{m,n}^S$. We assume that the speaker's observation of a given event is error-free. Then, the listener must reconstruct the event of the system based on the description transmitted by the speaker about $e_{m,n}^S$. We define $e_{m,n}^L$ as the reconstructed event by the listener at slot $n$ of episode $m$.
\vspace{-2mm}
\subsection{Events and Tasks}
As mentioned, the state of system is interpreted in the form of some events to the speaker. We define the set of the events, $\mathcal{E} =\big \{\mathcal{E}_{i}, \mathcal{E}_{\text{int}}, \mathcal{E}_{f}\big\}$, where $\mathcal{E}_{i}$ is the set of initial events, which indicate the start of new task. More precisely, if the observed event of the speaker is in $\mathcal{E}_{i}$, a new episode starts, and the speaker and listener must execute a new task. Thus, for each episode $m$, we have $e_{m,1}^S \in \mathcal{E}_{i}$. $\mathcal{E}_{f}$ is the set of final events capturing the end of a task. Therefore, for each episode $m$, we have $e_{m,E_m}^S \in \mathcal{E}_{f}$, where $E_m$ is the length of the executed task in episode $m$.
Finally, $\mathcal{E}_{\text{int}}$ is the set of intermediary events. Here, if $e_{m,n}^S \in \mathcal{E}_{\text{int}}$, then the speaker understands that the current ongoing task requires additional listener actions. 
To define the transition probability between different events, in two consecutive slots, we assume that event $e_{m,n+1}^S$ observed by the speaker in slot $n+1$ slot of episode $m$, is a function of event, $e_{m,n}^{S}$ and the action taken by the listener at slot $n$ of episode $m$ (see Section \ref{SubSec:Evolution} for more detail).

We can now define a \emph{task} based on different events. Based on the observed events, the speaker knows the state of each task and, thus, each task is composed of a chain of events from $\mathcal{E}$. Let $T_m \in \mathcal{T}$ be the task executed in episode $m$, where $\mathcal{T}$ is the set of all possible task types. Let $\mathcal{O}_k$ be the observed event chain of task type $k \in \mathcal{T}$, defined as follows:
\vspace{-1mm}
\begin{align}
\mathcal{O}_k = \Big\{\big( &e_{k,1}, e_{k,2}, \ldots, e_{k, L_k-1} , e_{k, L_k} \big) \Big | e_{k,1} \in \mathcal{E}_{i}, \notag\\ \big(& e_{k, 2}, \ldots, e_{k, L_k-1}\big) \in  \mathcal{E}_{\text{int}}, \, e_{k, L_k} \in \mathcal{E}_{f} \Big\}.
\label{Equation:Task_Def}
\end{align}

Due to the random dynamics of the environment, the length $L_k$ of a specific task, can vary in different episodes. Thus, the length of task $k$ is a random variable with probability mass function $f_k(L) = \text{Pr}(L_k = L)$, where $3 \leq L \leq L_{\text{max}}$. Here, the minimum length of each task type is $3$, because each task type has at least one event from each of $\mathcal{E}_{i}$, $\mathcal{E}_{\text{int}}$, and $\mathcal{E}_{f}$. We assume that each task's initial and final events are fixed. 
{\color{black}For example, consider an IoT system that needs to perform various control tasks. Each task begins with the appearance of a fixed state in the environment and ends with the appearance of another fixed state. Thus, different tasks types can be categorized based on their initial and final events.}
\vspace{-6mm}
\subsection{Beliefs as a Common Language}
\vspace{-1mm}
The speaker is responsible for perfectly describing the observed events to the listener in the most abstract manner. Hence, we consider the existence of a basic \emph{common language} between speaker and listener represented by a set $\mathcal{B}$ of $B$ beliefs. Each $b \in \mathcal{B}$ is a specific belief (feature in state-of-the-art) for describing the events. $\mathcal{B}$ is an input to the problem known to the speaker and listener. However, the set of beliefs can vary for different applications. Intuitively, the set of beliefs for each application can be achieved by conducting a feature selection procedure. We make three assumptions regarding the belief-based common language: a) \emph{Consensuality:} The speaker and listener have the same perception about each member of the belief set, b) \emph{Comprehensivity:} The belief set is comprehensive in a way that each event can be perfectly described by a subset of it, and c) \emph{Abstraction:} The belief set is essentially a set of abstract features of events in a way that reduces the transmission cost between the speaker and listener.

In our system, the speaker and listener have the same perception about the mapping between beliefs and the observed events. Therefore, using all of the beliefs in $\mathcal{B}$ is a perfect descriptor of each event. However, we assume that, for each event, there is a small subset of $\mathcal{B}$ which can perfectly describe the event, and the rest of the beliefs are unnecessary. Thus, using all of the beliefs for describing each event is a naive policy. Our objective is to find the most abstract subset of $\mathcal{B}$ for perfectly describing each event which directly impacts the task execution by the listener. We consider the most abstract description of each event as a \emph{semantic realization} of the event, that contains sufficient information for the listener's correct decision-making for the ongoing task. Although each belief can have information about each event, some beliefs may not be helpful for the decision-making of the listener and can be discarded. Hereinafter, when we state finding the perfect description of an event, we imply describing the event using the most abstract subset of $\mathcal{B}$, i.e., the \emph{semantics}.

There could be multiple subsets of $\mathcal{B}$ that perfectly describe each event. Hence, we define $\mathcal{B}_{e_j}^P = \big\{\mathcal{B}_l \big| \mathcal{B}_l \subset \mathcal{B}, \mathcal{B}_l \text{ is perfect description of } e_j \big\}$ as the set of all possible subsets of $\mathcal{B}$ that perfectly describe $e_j$. To find the most abstract description of each event, we define two major decision metrics. First, the \emph{transmission cost} of each subset of $\mathcal{B}$. Indeed, for transmitting each $b \in \mathcal{B}$ the speaker will incur a cost of $C_b$ that is belief-dependent.
Second, we consider the number of transmitted beliefs captured by the size of the used description.
We combine these two metrics to determine the abstract description of each event. 
We define $\mathcal{B}_{m,n}^S \in \mathcal{B}$ and $C_{m,n}^S$ as the description of the speaker and its metric in slot $n$ of episode $m$, respectively. 
The value of $C_{m,n}^S$ can be computed as follows:
\vspace{-1mm}
\begin{align}
C_{m,n}^S &= \alpha \textstyle\sum_{b=1}^{B}{C_b \times x_{m,n,b}^S} + (1 - \alpha)  \textstyle\sum_{b=1}^{B}{x_{m,n,b}^S}, \label{Equation:Belief_Cost}
\end{align}
where $x_{m,n,b}^S$ is a binary variable that indicates whether belief $b$ is used in $\mathcal{B}_{m,n}^S$ or not. $\alpha$ is a design parameter for capturing the importance of each metric. We normalize the values of $C_b$ to bring them within the range of the cardinality metrics.When the speaker is aware of the subsets which perfectly describe an event, it only needs to determine the optimum subset using \eqref{Equation:Belief_Cost}. However, this is not realistic and, thus, we assume the speaker to be unaware of the importance of each belief in describing the events, for the decision-making of the listener. 
Here, the speaker should gradually learn the importance of each belief for various events by executing different tasks. Specifically, the speaker can gain this knowledge by using the different subsets of $\mathcal{B}$ for the description of the events and considering their effect on the task execution process.
\vspace{-2mm}
\subsection{System State Evolution}
\label{SubSec:Evolution}
\vspace{-1mm}
The state of the system at time slot $n$ of episode $m$ is represented by $e_{m,n}^S$ observed by the speaker.
Also, we use $e_{m,n}^L$ to denote the perceived event of the listener at time slot $n$ of episode $m$. The listener builds $e_{m,n}^L$ based on the transmitted belief of the speaker, $\mathcal{B}_{m,n}^S$.
The transition from $e_{m,n}^S$ to $e_{m,n+1}^S$ is dependent on $e_{m,n}^L$, which is function of $\mathcal{B}_{m,n}^S$. We define $p_{j,j^{\prime}}$ as the transition probability from $e_{m,n}^S = e_j$ to $e_{m,n+1}^S = e_{j^{\prime}}$, as follows:
\vspace{-2mm}
\begin{align}
p_{j,j^{\prime}} &= \text{Pr} \big (e_{m,n}^S = e_j \to e_{m,n+1}^S = e_{j^{\prime}} \big| \mathcal{B}_{m,n}^S  \big ) \notag \\
&=\begin{cases} \boldsymbol{P}_{j,j^{\prime}} & \mathcal{B}_{m,n}^S \in \mathcal{B}_{e_j}^{P},   \\ \boldsymbol{\tilde{P}}_{j,j^{\prime}} & \mathcal{B}_{m,n}^S \notin \mathcal{B}_{e_j}^{P},\end{cases}
\label{Equation:Transition_Prob}
\end{align}
where the case $\mathcal{B}_{m,n}^S \in \mathcal{B}_{e_j}^{P}$, is a scenario in which the transmitted belief of the speaker perfectly describes the observed event, and $\boldsymbol{P}$ is the transition probability matrix when the listener takes proper action regarding the observed event. The second case captures a scenario in which the transmitted belief of the speaker is not a perfect descriptor of the event. Thus, the listener cannot take the right action. Also, $\boldsymbol{\tilde{P}}$ is the transition probability matrix of the system when the taken action of the listener is not proper for the observed event. We expect $\boldsymbol{P}$ to be a sparse matrix, while $\boldsymbol{\tilde{P}}$ will be a random matrix that has more nonzero elements than $\boldsymbol{P}$. Also, we have $e_{m,n}^L = e_{m,n}^S$ for $\mathcal{B}_{m,n}^S \in \mathcal{B}^P_{e_j}$ and $e_{m,n}^L = e_j \in \{\mathcal{E} \setminus e_{m,n}^S\}$ when $\mathcal{B}_{m,n}^S \notin \mathcal{B}^P_{e_j}$, where $\{\mathcal{E} \setminus e_{m,n}^S\}$ is the set of all events except $e_{m,n}^S$.
\vspace{-2mm}
\subsection{Task Execution Procedure}
\label{Subsec:Task_Execution}
\vspace{-1mm}
If the task executed during episode $m$, is of type $k$, then $e_{m,1}^S = e_{k,1}$ and $e_{m,E_m}^S = e_{k, L_k}$, for the initial and final events of episode $m$. The observed intermediary events depend on the performance of the speaker and listener as well as the randomness of the environment, modeled by $f_k$. If the speaker perfectly describes each observed event, then the observed intermediary events only depend on $f_k$, and thus, the length of episode $m$ will be determined by the mean of $f_k$.
Although using all of the beliefs in $\mathcal{B}$ is a perfect descriptor of each event and minimizes task execution time, it is not optimal in terms of transmission cost.
Thus, the speaker and listener's objective is to find the most abstract description of each event, which simultaneously minimizes the task transmission cost and task execution time. Let $C_m^S= \sum_{n=1}^{E_m}{C_{m,n}^S}$ be the transmission cost of the executed task in episode $m$. Thus, we have:
\vspace{-2mm}
\begin{align}
\hspace{-1.95mm}C_{m}^S= \alpha \sum_{n=1}^{E_m}{ \sum_{b=1}^{B}{C_b \times x_{m,n,b}^S}} +(1 - \alpha)  \sum_{n=1}^{E_m}{\sum_{b=1}^{B}{x_{m,n,b}^S}}  \label{Equation:Task_Cost}. 
\end{align}

The transmission cost of a task execution depends on the number of time slots, $E_m$, the task type, and the number of beliefs used for describing the observed events. The randomness of the executed task plays a key role in determining $E_m$.
\vspace{-6mm}
\subsection{Optimization of Semantic Communications}
\vspace{-1mm}
The speaker's goal is to simultaneously minimize the belief transmission cost defined in \eqref{Equation:Task_Cost}, and the task execution time. The required time for executing the task of episode $m$ is essentially the length $E_m$ of this episode. Hence, we should consider the length of each episode in the optimization. The optimization problem must balance the tradeoff between minimizing the task execution time and belief transmission cost, as follows:
\vspace{-3mm}
\begin{subequations}
\begin{align}
\hspace{-2mm}\min_{x_{m,n,b}^{S} \in \{0,1\}} & \,\Big [\lim_{M \to \infty} \sum_{m=1}^{M}{\Big(\delta \times C_m^S  + (1 - \delta) \times  E_m \Big)} \Big ], \label{eq:Objective}\\
 \text{s. t.} &\; 1 \leq \textstyle\sum_{b=1}^{B}{x_{m,n,b}^S \leq \frac{B}{2}}, 
\end{align}
\label{eq:Optimization}
\end{subequations}
\hspace{-1.6mm}where $0 \leq \delta \leq 1$ is a parameter to balance the discussed tradeoff. $C_m^S$ is function of $x_{m,n,b}^S$ according to \eqref{Equation:Task_Cost}, but, for simplicity, we do not show this dependence in \eqref{eq:Objective}.
The first term in \eqref{eq:Objective} is required to find the most abstract description of each event, and the second term is critical to minimizing task execution time. Thus, the objective of \eqref{eq:Optimization} is to find an optimal description of each event so as to minimize the belief transmission cost and task execution time in an infinite horizon. The first constraint restricts the number of used beliefs in $\mathcal{B}_{m,n}^S$.

Problem \eqref{eq:Optimization} cannot be solved using standard optimization techniques and Markov decision process (MDP), because the dynamic transitions between events (matrices $\boldsymbol{\tilde{P}}$ and $\boldsymbol{P}$), which are unknown to our system. Here, it is apropos to use model-free RL \cite{sutton2018reinforcement}. As the RL agent, the speaker wants to gradually learn the importance of each belief for each event by experiencing various tasks to find each event's abstract and perfect description. However, using traditional RL could yield long training with poor performance during training because of two reasons: 1) The size of the action space is $2^{B}$, which is exponential, and 2) In our model, the agent can evaluate each action in the task execution procedure. Thus, the reward signal depends on the task execution, resulting in a sparse reward signal, which means that the agent only receives a positive reward signal when he completes the task. Furthermore, the randomness of the environment, which implies randomness in the task execution time, is a major challenge. Due to these factors, when training traditional RL, the task execution time can be significantly longer than the expectation determined by $f_k$, which is unacceptable for goal-oriented communication services such as control tasks in an IoT network. These challenges motivate us to solve this problem using a curriculum framework based on RL.

\vspace{-1mm}
\section{Proposed Curriculum Learning (CL) Method}
\label{Section: Proposed_Method}
\vspace{-1mm}
Our main goal is to learn the abstract description of each event to minimize the task execution time and transmission cost. The speaker can gain this knowledge by experiencing various tasks with different events.
However, similar to the case of traditional RL, the random exploration step increases training time and could yield a poor performance during training.
In such problems, using a curriculum framework can be a solution \cite{narvekar2020curriculum}. The main idea behind CL is to design simple tasks based on the main problem, perform these simple tasks, and use their gained experience to solve the main, more complicated. In our model, the action space, which is the different subsets of $\mathcal{B}$, is the main challenge. Thus, we can design the simple tasks based on a reduction of the action space. Precisely, to leverage CL in our problem, we can evaluate the various subsets of $\mathcal{B}$ with different cardinalities, separately. For this purpose, we propose a top-down CL framework.

\vspace{-2mm}
\subsection{Top-Down Curriculum Learning}
The proposed top-down CL approach begins with a comprehensive subset of $\mathcal{B}$, as a perfect descriptor of all events, and then, it performs event-specific pruning to find the most abstract description of each event. We assume that there is a perfect descriptor $\mathcal{B}^{\text{comp}} = \big\{\bigcup_{e_j \in \mathcal{E}}{\mathcal{B}^{\text{opt}}_{e_j}}\big| \, \mathcal{B}^{\text{opt}}_{e_j} \in \mathcal{B}^S_{e_j}, \, \big|B^{\text{opt}}_{e_j}\big| = \min_{\mathcal{B}_l \in \mathcal{B}^S_{e_j}}{\big| \mathcal{B}_l \big|} \big\}$ for all events. $\mathcal{B}^{\text{opt}}_{e_j}$ is the most abstract and perfect descriptor of the event $e_j$. Finding $\mathcal{B}^{\text{comp}}$ can be time-consuming because it requires finding the abstract description of each event. Meanwhile, $\mathcal{B}^{\text{comp}}$ is a large set relative to the scale of $\mathcal{B}$. Hence, we can consider the most comprehensive option for $\mathcal{B}^{\text{comp}}$, which is $\mathcal{B}$, and start pruning.

The pruning procedure is the main part of our method, whereby $\mathcal{B}^{\text{comp}}$ is pruned for each event in a way that the most abstract and perfect descriptor of each event is determined. Here, the objective is minimizing the task execution time and the belief transmission cost. As such, we propose 
{\emph{a linear sequence-based and task-level CL}. The term ``task'' is used here in the context of the CL paradigm, and it is not related to our system's task. We now define a \emph{curriculum} as an ordered list of tasks $[z_1, z_2, \ldots, z_l]$ \cite{narvekar2020curriculum} with $z_l$ being the task of finding the subsets of $\mathcal{B}^{\text{comp}}$ with cardinality $l$, that can be pruned for each event. Precisely, in $z_l$, for each event $e_j$, we want to find all possible subsets of $\mathcal{B}^{\text{comp}}$ with cardinality $l$, which can be individually eliminated from $\mathcal{B}^{\text{comp}}$ while maintaining a perfect description of $e_j$. Let $\mathcal{B}^{\text{prun}}_{e_j, z_l}$ be the set of these subsets for event $e_j$, in step $z_l$. Each member of $\mathcal{B}^{\text{prun}}_{e_j, z_l}$ is individually unnecessary for a perfect description of $e_j$ and thus it can be pruned from $\mathcal{B}^{\text{comp}}$ for describing $e_j$. To reduce the complexity of this pruning, we design a curriculum framework in which the agent starts from $1$-length belief pruning. Using the gained experience at each step, the agent gradually investigates more difficult pruning. We define $\mathcal{B}^{\text{prun}}_{e_j, z_1} = \big\{b_i  | b_i \in \mathcal{B}^{\text{comp}}, \; \mathcal{B}^{\text{comp}} \setminus b_i \in \mathcal{B}^P_{e_j}\big\}$ as the output of the first step of the curriculum for $e_j$.

We can define CL task $z_l$ for $e_j$ as the task of finding all subsets of $\mathcal{B}^{\text{comp}}$ with cardinality $l$, which can be individually eliminated from $\mathcal{B}^{\text{comp}}$ while maintaining a perfect description of $e_j$, as follows:
\vspace{-2.5mm}
\begin{align}
&\mathcal{B}^{\text{prun}}_{e_j,z_l} = \Big\{\big\{b_{i_1}, b_{i_2}, \ldots, b_{i_l}\big\} \big | b_{i_1} \in \mathcal{B}^{\text{prun}}_{e_j,z_1} , \; \label{Equation:Curriculum_Learning_3}  \\
\big\{&b_{i_2}, \ldots, b_{i_l}\big\} \in \mathcal{B}^{\text{prun}}_{e_j,z_{l-1}}, \,
\mathcal{B}^{\text{comp}} \setminus \big\{b_{i_1}, \ldots, b_{i_l}\big\} \in \mathcal{B}^P_{e_j})\Big\} \notag
\end{align}

The gained experiences at each step are directly used for initializing the next CL step. This is why we categorize our proposed method as CL.
One of the main challenge in a CL method is how to transfer the gained experiences in each step to the next step to simplify it.
The transfer is needed to extract and pass on reusable the experience acquired from one step to the next. The main transfer methods are low-level knowledge such as an entire policy or directly initializing the agent in the next step \cite{narvekar2020curriculum}. The transfer method that we adopt directly initializes the agent in the next step, as seen in \eqref{Equation:Curriculum_Learning_3}.
\subsection{Learning at each CL Step}
\label{Section: Method_Each_Step}
\looseness=-1
Now, we introduce an RL method using Q-Learning for solving each step of the CL. We use RL here because the agent is unaware of the transition matrices $\boldsymbol{P}$ and $\boldsymbol{\tilde{P}}$ of the events. 

\subsubsection{RL for the first CL step}
\label{Sec:Framework_First_Step}
We seek an RL method capable of determining $\mathcal{B}^{\text{prun}}_{e_j, z_1}$ for each $e_j \in \mathcal{E}$. The speaker is our RL agent. Here, our RL problem can be defined using a tuple $(\Omega^S_{z_1}, \Omega^A_{z_1}, \Omega^R_{z_1})$, where $\Omega^S_{z_1}$ is the state space, $\Omega^A_{z_1}$ is the action space, and $\Omega^R_{z_1}$ is the reward space. We define the state space in a way that satisfies the Markovian property. We define the \emph{state space} of the first step of CL, $\Omega^S_{z_1}$, as the set of all possible events that the agent can observe, $\Omega^S_{z_1} = \big\{e_j| e_j \in \{\mathcal{E}_{i},  \mathcal{E}_{\text{int}}, \mathcal{E}_{f}\}\big\}$. This state space includes all of the initial, intermediary, and final events. We define $\lambda_{m,n} \in \Omega^S_{z_1}$ the state of the agent at time slot $n$ of episode $m$, which represents the observed event of the speaker, $e_{m,n}^S$.

In the first step of CL, for each $e_j$, we want to find all possible subsets of $\mathcal{B}^{\text{comp}}$ with cardinality $1$ that can be pruned. Thus, we define the first step's \emph{action space}, as all possible subsets of $\mathcal{B}^{\text{comp}}$ with cardinality of $1$, as $\Omega^A_{z_1} = \big\{b_i | b_i \in \mathcal{B} \big\}$. We define $a_{m,n}$ the action of the agent in time slot $n$ of episode $m$. Now, the used description of the speaker in time slot $n$ of episode $m$, $\mathcal{B}_{m,n}^S$, will be:
\vspace{-3mm}
\begin{align}
 \mathcal{B}_{m,n}^S = \Big\{ \mathcal{B}^{\text{comp}} \setminus a_{m,n} \big |  a_{m,n} \in \Omega^A_{z_1} \Big\}.
\label{Equation:Belief_Action_Relation}
\end{align}

Here, the speaker will eliminate $a_{m,n}$ from $\mathcal{B}^{\text{comp}}$, and use the remaining beliefs for the description of the observed event.

The \emph{reward function}, $R_{z_1}$, defined as $R_{z_1}:\Omega^S_{z_1} \times \Omega^A_{z_1} \to \Omega^R_{z_1}$, captures the expected immediate reward gained by the agent for taking each action in each state. We define $R_{z_1}(\lambda_{m,n}, a_{m,n})$ as the gained reward of taking action $a_{m,n} \in \Omega^A_{z_1}$ at state $\lambda_{m,n} \in \Omega^S_{z_1}$. For defining $R_{z_1}$, we must consider the objectives of the system, which include minimizing the task execution time and belief transmission cost. The effect of an action on the task execution procedure can be determined, based on the current observed event and the next event. Therefore, we have:
\vspace{-3mm}
\begin{align}
R_{z_1}(\lambda_{m,n}, a_{m,n}) = \begin{cases} - C_{m,n}^S + R_k& \lambda_{m,n+1} \in \mathcal{E}_{f},   \\ 
								      -C_{m,n}^S - C_k & \lambda_{m,n+1} \in \mathcal{E}_{i}, \\
								      -C_{m,n}^S & \lambda_{m,n+1} \in \mathcal{E}_{\text{int}}  .\end{cases}
\label{Equation:Reward_Definition_First_Step}								      
\end{align}

\looseness=-1
Here, $C_{m,n}^S$ is the transmission cost of the used description in slot $n$ of episode $m$, defined in \eqref{Equation:Belief_Cost}. $k$ is the type of executed task in episode $m$. $R_k$ and $C_k$ are the reward and cost of task execution and delay for type $k$, respectively. The first case in \eqref{Equation:Reward_Definition_First_Step} is the scenario in which the next observed event, $\lambda_{m,n+1}$ is a final event, and thus, the ongoing task in episode $m$ is completed. We consider a reward $R_k$ to encourage the agent to take such actions in each event. The second case is when the next event is an initial event. The initial events can only be observed at the first time slot of each episode. Thus, if we have $\lambda_{m,n+1} \in \mathcal{E}_{i}$ for $n \geq 1$, then the action taken at time slot $n$ is delaying the task execution. Hence, we consider a cost $C_k$ for such actions. The last case captures the scenario in which the next observed event is an intermediary event, where we only consider the transmission cost of the taken action.

\subsubsection{RL for Step $l$ of the CL}
\label{Sec:Framework_L_Step}
Now, we want to define $\Omega^S_{z_l}$, $\Omega^A_{z_l}$, and $\Omega^R_{z_l}$ for the state space, action space, and reward space of step $l$ of the CL method. The state space and reward space of step $l$ are defined as the same as the first step. However, the action space definition for step $l \geq 2$ differs from the first step. Here, we use the gained experiences in the previous steps to initialize the action space. We also must define a specific action space $\Omega^A_{z_l, e_j}$ for each event $e_j$ in step $l \geq 2$. We define $\Omega^A_{z_l,e_j}$, based on the output of the first and $(l-1)^\text{th}$ CL steps, captured by $\mathcal{B}^{\text{prun}}_{e_j, z_1}$, and $\mathcal{B}^{\text{prun}}_{e_j, z_{l-1}}$. Therefore, we have:
\vspace{-2mm}
\begin{align}
\Omega^A_{z_l,e_j} = \Big\{ \big\{&b_{i_1}, b_{i_2}, \ldots, b_{i_l}\big\} \big | b_{i_1} \in \mathcal{B}^{\text{prun}}_{e_j,z_1} , \; \notag \\
\big\{&b_{i_2}, b_{i_3}, \ldots, b_{i_l}\big\} \in \mathcal{B}^{\text{prun}}_{e_j,z_{l-1}} \Big\}.
\label{Equation:Action_Definition_L_Step}
\end{align}

Each element of $\mathcal{B}^{\text{prun}}_{e_j,z_1}$ and $\mathcal{B}^{\text{prun}}_{e_j,z_{l-1}}$, can be separately eliminated from $\mathcal{B}^{\text{comp}}$ while maintaining a perfect description of $e_j$. Hence, the candidates for pruning at step $l$ of CL are the combinations of the elements of these sets. The computation of $\mathcal{B}_{m,n}^S$ in step $l$ of the CL is similar to the first step in \eqref{Equation:Belief_Action_Relation}.
\vspace{-2mm}
\subsection{Proposed Top-Down CL Algorithm}
\vspace{-1mm}
In Algorithm \ref{Algo_CL}, we summarize the proposed top-down CL algorithm for solving the introduced optimization problem. In step \ref{Algo_CL_Initialization1}, we initialize the input variables.
In \ref{Algo_CL_L1_Start}-\ref{Algo_CL_L1_End}, the first step of CL method is executed. In \ref{Algo_CL_L2_Start}-\ref{Algo_CL_L2_End}, the steps $l \geq 2$ of CL method are executed. 
The main difference between the first step and the other steps is in the action space. In the first step, we have a unique action space for all events. However, in step $l \geq 2$, we have to define a specific action space for each event, $e_j$. In steps \ref{Algo_CL_Event_Check_Start}-\ref{Algo_CL_Event_Check_End}, the algorithm checks whether the pruning procedure for an event has terminated or not. If the pruning has finished for $e_j \in \mathcal{E}$, then the algorithm determines the optimal description of this event, in step \ref{Algo_CL_Descript_Opt}. In \ref{Algo_CL_Termin_Start}-\ref{Algo_CL_Termin_End}, the algorithm checks the termination of the pruning for all of the events to end the CL steps. In step \ref{Algo_CL_Output_Cal}, the outputs of the algorithm are determined.

\begin{algorithm}[t]
	\label{Algo_CL}
	\small
		
	\caption{\small{Top-Down Curriculum Learning}}
	
\footnotesize{	\textbf{Inputs:} $\mathcal{E}, \mathcal{B},  C_i\text{: The cost of using belief }i$, \text{and }$\mathcal{B}^{\text{comp}} = \mathcal{B}$. \label{Algo_CL_Initialization1}\\
	\For{($l =1:B-1$)} 
	{\label{Algo_CL_CL_Start}
		\If{$ l = 1$}
		{\label{Algo_CL_L1_Start}
			$\Omega^S_{z_1} = \mathcal{E}, \Omega^A_{z_1} = \mathcal{B}$.\\
			Perform RL by $\Omega^S_{z_1}$, $\Omega^A_{z_1}$, and reward function of \eqref{Equation:Reward_Definition_First_Step}.\\
			Calculate $\mathcal{B}^{\text{prun}}_{e_j, z_1}$ for each $e_j$, using $Q$ values.\\
			\vspace{-1.2mm}
		}\label{Algo_CL_L1_End}
		\vspace{-0.4mm}
		\If{$ l > 1$}
		{\label{Algo_CL_L2_Start}
			$\Omega^S_{z_l} = \mathcal{E}$ and compute $\Omega^A_{z_l,e_j}$ using \eqref{Equation:Action_Definition_L_Step}.\\
		        Perform RL by $\Omega^S_{z_l}$, $\Omega^A_{z_l,e_j}$, and reward function of \eqref{Equation:Reward_Definition_First_Step}.\\
		        Calculate $\mathcal{B}^{\text{prun}}_{e_j, z_l}$ for each $e_j$, using $Q$ values.\\
		        \vspace{-1.2mm}
		}\label{Algo_CL_L2_End}
		\vspace{-0.6mm}
		\For{($e_j \in \mathcal{E}$)}
		{\label{Algo_CL_Event_Check_Start}
			\If{$\mathcal{B}^{\text{prun}}_{e_j, z_l} = \emptyset$}
			{
				$\mathcal{B}^{\text{opt}}_{e_j} = \mathcal{B}^{comp} \setminus \mathcal{B}^{\text{prun}}_{e_j, z_{l-1}}$. \label{Algo_CL_Descript_Opt}
				\vspace{-2mm}
			}\vspace{-1.2mm}
		}\label{Algo_CL_Event_Check_End}
		\vspace{-0.4mm}	
		\If{$\mathcal{B}^{\text{prun}}_{e_j, z_l} = \emptyset, \forall e_j \in \mathcal{E}$}
		{\label{Algo_CL_Termin_Start}
			$\mathbf{Break}$.
			\vspace{-1.2mm}
		}\label{Algo_CL_Termin_End}
		\vspace{-0.4mm}
	}\label{Algo_CL_CL_End}
	\vspace{-0.4mm}
	\textbf{Outputs:} $\mathcal{B}^{\text{opt}}_{e_j}, \forall e_j \in \mathcal{E}$.\label{Algo_CL_Output_Cal}
	\vspace{-0.4mm}}
\end{algorithm}


\begin{figure}[!t]
\centering
\vspace{-3mm}
	\includegraphics[width=0.35\textwidth]{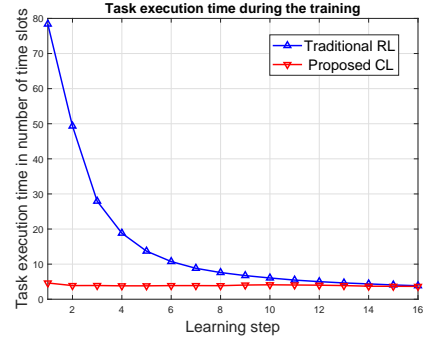}
	\vspace{-3mm}
	\caption{{Comparison of the proposed CL with traditional RL in terms of the task execution time.}}
	\label{Task_Time_Comparison} 
	\vspace{-5mm}
\end{figure}

\begin{figure}[!t]
\centering
	\includegraphics[width=0.35\textwidth]{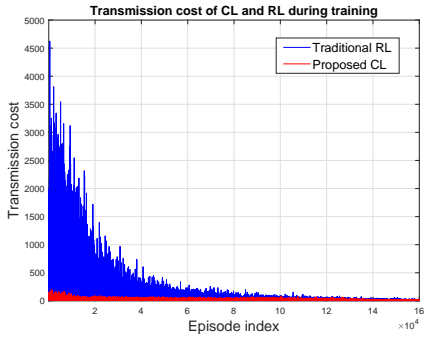}
	\vspace{-3mm}
	\caption{{Comparison of the proposed CL with traditional RL in terms of the belief transmission cost for each task execution.}}
	\label{Belief_Transmission_Comparison} 
	\vspace{-3mm}
\end{figure}

\begin{figure}[!t]
\centering
\begin{subfigure}[\scriptsize{Transmission cost per task in different pruning steps.}]{
\includegraphics[width=0.34\textwidth]{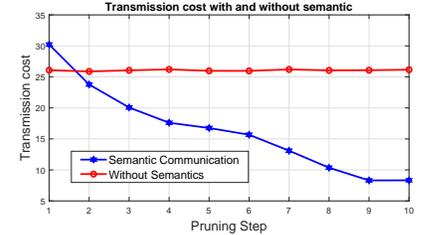} 
\vspace{-3mm}
\label{Non_Semantic_Cost}
}
\vspace{-3mm}
\end{subfigure}

\begin{subfigure}[\scriptsize{Task execution time in different pruning steps.}]{
\includegraphics[width=0.34\textwidth]{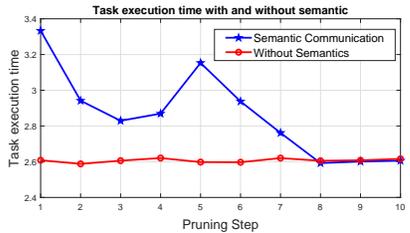}
\label{Non_Semantic_Time}
}
\end{subfigure}

\vspace{-2mm}
\caption{Benefits of using semantic communications with the proposed CL method.}
\label{Non_Semantic} 
\vspace{-6mm}
\end{figure}

\section{Simulation Results and Analysis}
\label{Section: Numerical_Results}
\looseness=-1
For our simulations, we consider $10$ beliefs and $10$ tasks. The cost of using different beliefs is randomly generated between $1$ and $2$. We consider $60$ events, including the initial, intermediary, and final events.
We assume that the abstract and perfect description of each event can have $2$ to $4$ beliefs.
The minimum and maximum number of events in each task are considered to be $3$ and $6$. 
The event transition matrices $\boldsymbol{P}$ and $\boldsymbol{\tilde{P}}$ are of size $60 \times 60$. These matrices are generated according to the chain of events of each task.
For comparison purposes, we use a traditional RL solution that performs random exploration between all subsets of $\mathcal{B}$ to solve the optimization in \eqref{eq:Optimization}.


\looseness=-1
Fig. \ref{Task_Time_Comparison} shows the task execution time resulting from both CL and RL. We averaged the task execution time over the last $10,000$ of episodes for each method. Fig. \ref{Task_Time_Comparison} shows that the task execution time for the CL method is approximately constant (compared to traditional RL) during training because the CL method is based on gradually pruning the unnecessary beliefs without significantly affecting the task execution time. This figure shows that the proposed CL method yields a three-fold improvement in the task execution time compared to traditional RL. Fig. \ref{Task_Time_Comparison} includes $160,000$ episodes for both methods. 
\looseness=-1

In Fig. \ref{Belief_Transmission_Comparison}, the belief transmission cost for each executed task during training is presented. Fig. \ref{Belief_Transmission_Comparison} shows that the transmission cost of both methods gradually decreases by finding the abstract description of the events. However, due to the random exploration in RL, the RL method experiences large values for the transmission cost of the executed task during training. In contrast, the fluctuation in the transmission cost values of the CL method is reasonable during training. Precisely, on average, the CL method yields a two-fold improvement in the transmission cost during training, compared to traditional RL.

\looseness=-1
Fig. \ref{Non_Semantic} evaluates the performance of the introduced semantic goal-oriented model compared to a case without semantics. In the non-semantic scenario, the speaker uses all of the beliefs in $\mathcal{B}$ for describing each event in each time slot. The x-axis indicates the different pruning steps. Figs. \ref{Non_Semantic_Cost}-\ref{Non_Semantic_Time} show that at the beginning step of the pruning, the transmission cost and task execution time of our model is higher than the non-semantic. However, by pruning the unnecessary beliefs one by one, the transmission cost decreases and at the end of pruning yields a two-fold improvement compared to non-semantic, as seen in Fig. \ref{Non_Semantic_Cost}. In addition, the task execution time is also reduced by gradually pruning the unnecessary beliefs, as seen in Fig. \ref{Non_Semantic_Time}. The increase of the task execution time at pruning step $5$  stems from the increase in the size of $\mathcal{B}^{\text{prun}}_{e_j,z_{l}}$. In particular, for our simulation setup $\mathcal{B}^{\text{prun}}_{e_j,z_{l}}$ for $l=5$ has the greatest cardinality, thereby resulting in more imperfect descriptions of events at step $5$ of pruning and increasing the task execution time. However, subsequently, the size of $\mathcal{B}^{\text{prun}}_{e_j,z_{l}}$ decreases, and the task execution time of our model is reduced and converges to non-semantic.

\vspace{-1mm}
\section{Conclusion}
\label{Section: Conclusion}
	In this paper, we have studied the problem of goal-oriented semantic communications. We have defined a new model that allows a speaker and a listener to use a common language, called beliefs, to execute system tasks based on observed environmental events. Then, we have introduced an optimization problem to find the abstract and perfect description of each event to minimize the task execution time and the belief transmission cost. For solving the introduced optimization problem, we have proposed a novel CL framework, which determines the optimum description of each event by gradually eliminating the unnecessary beliefs for each event. Simulation results show how the proposed CL method significantly improves the task execution time and reduces transmission costs compared to traditional RL and classical non-semantic models.

\vspace{-1mm}
\bibliographystyle{IEEEtran}
\bibliography{ref}

\begin{thebibliography}{10}
\providecommand{\url}[1]{#1}
\csname url@samestyle\endcsname
\providecommand{\newblock}{\relax}
\providecommand{\bibinfo}[2]{#2}
\providecommand{\BIBentrySTDinterwordspacing}{\spaceskip=0pt\relax}
\providecommand{\BIBentryALTinterwordstretchfactor}{4}
\providecommand{\BIBentryALTinterwordspacing}{\spaceskip=\fontdimen2\font plus
\BIBentryALTinterwordstretchfactor\fontdimen3\font minus
  \fontdimen4\font\relax}
\providecommand{\BIBforeignlanguage}[2]{{%
\expandafter\ifx\csname l@#1\endcsname\relax
\typeout{** WARNING: IEEEtran.bst: No hyphenation pattern has been}%
\typeout{** loaded for the language `#1'. Using the pattern for}%
\typeout{** the default language instead.}%
\else
\language=\csname l@#1\endcsname
\fi
#2}}
\providecommand{\BIBdecl}{\relax}
\BIBdecl

\bibitem{strinati20216g}
E.~C. Strinati and S.~Barbarossa, ``6{G} networks: Beyond {S}hannon towards
  semantic and goal-oriented communications,'' \emph{Computer Networks}, vol.
  190, 2021.

\bibitem{xie2021deep}
H.~Xie~et al., ``Deep learning enabled semantic communication systems,''
  \emph{IEEE Transactions on Signal Processing}, vol.~69, pp. 2663--2675, 2021.

\bibitem{xie2020lite}
H.~Xie and Z.~Qin, ``A lite distributed semantic communication system for
  internet of things,'' \emph{IEEE Journal on Selected Areas in
  Communications}, vol.~39, no.~1, pp. 142--153, Nov. 2020.

\bibitem{guler2018semantic}
B.~G{\"u}ler, A.~Yener, and A.~Swami, ``The semantic communication game,''
  \emph{IEEE Transactions on Cognitive Communications and Networking}, vol.~4,
  no.~4, pp. 787--802, Sep. 2018.

\bibitem{lu2021reinforcement}
K.~Lu~et al., ``Reinforcement learning-powered semantic communication via
  semantic similarity,'' \emph{arXiv preprint arXiv:2108.12121}, 2021.

\bibitem{yun2021attention}
W.~J. Yun~et al., ``Attention-based reinforcement learning for real-time
  {U}{A}{V} semantic communication,'' \emph{arXiv preprint arXiv:2105.10716},
  2021.

\bibitem{tucker2021emergent}
M.~Tucker~et al., ``Emergent discrete communication in semanticspaces,''
  \emph{arXiv preprint arXiv:2108.01828}, 2021.

\bibitem{maarala2016semantic}
A.~I. Maarala, X.~Su, and J.~Riekki, ``Semantic reasoning for context-aware
  internet of things applications,'' \emph{IEEE Internet of Things Journal},
  vol.~4, no.~2, pp. 461--473, Jul. 2016.

\bibitem{seo2021semantics}
H.~Seo~et al., ``Semantics-native communication with contextual reasoning,''
  \emph{arXiv preprint arXiv:2108.05681}, 2021.

\bibitem{sutton2018reinforcement}
R.~S. Sutton and A.~G. Barto, \emph{Reinforcement learning: An
  introduction}.\hskip 1em plus 0.5em minus 0.4em\relax MIT press, 2018.

\bibitem{narvekar2020curriculum}
S.~Narvekar~et al., ``Curriculum learning for reinforcement learning domains: A
  framework and survey,'' \emph{arXiv preprint arXiv:2003.04960}, 2020.

\end{thebibliography}
\end{document}